\documentclass[lettersize,journal]{IEEEtran}
\usepackage{amsmath,amsfonts}
\usepackage{algorithmic}
\usepackage{algorithm}
\usepackage{array}
\usepackage[caption=false,font=normalsize,labelfont=sf,textfont=sf]{subfig}
\usepackage{textcomp}
\usepackage{stfloats}
\usepackage{url}
\usepackage{verbatim}
\usepackage{graphicx}
\usepackage{cite}
\begin{document}

\title{Edge Caching Optimization with PPO and Transfer Learning for Dynamic Environments}

\author{Farnaz Niknia, 
Ping Wang,~\IEEEmembership{Fellow,IEEE}

\thanks{Farnaz Niknia and Ping Wang are with the Lassonde School of Engineering at York University, Toronto, ON, M3J 1P3, Canada (email: fniknia@yorku.ca, ping.wang@lassonde.yorku.ca)}

}

\maketitle

\begin{abstract}
This paper addresses the challenge of edge caching in dynamic environments, where rising traffic loads strain backhaul links and core networks. We propose a Proximal Policy Optimization (PPO)-based caching strategy that fully incorporates key file attributes such as size, lifetime, importance, and popularity, while also considering random file request arrivals, reflecting more realistic edge caching scenarios. In dynamic environments, changes such as shifts in content popularity and variations in request rates frequently occur, making previously learned policies less effective as they were optimized for earlier conditions. Without adaptation, caching efficiency and response times can degrade. While learning a new policy from scratch in a new environment is an option, it is highly inefficient and computationally expensive. Thus, adapting an existing policy to these changes is critical. To address this, we develop a mechanism that detects changes in content popularity and request rates, ensuring timely adjustments to the caching strategy. We also propose a transfer learning-based PPO algorithm that accelerates convergence in new environments by leveraging prior knowledge. Simulation results demonstrate the significant effectiveness of our approach, outperforming a recent Deep Reinforcement Learning (DRL)-based method. 
\end{abstract}


\section{Introduction}

\IEEEPARstart{W}{ith} the rapid growth of mobile applications and services, the demand for data has surged, placing significant pressure on backhaul links and core networks. The exponential rise in video streaming, cloud services, and Internet of Things (IoT) devices has led to an overwhelming number of content requests, often from centralized data centers. To address this, edge caching frequently requested data closer to users, has become a key strategy for reducing network congestion and minimizing transmission delays. By caching popular content locally, edge caching not only alleviates the load on core networks but also improves the overall user experience by reducing latency.

Edge caching faces significant challenges in dynamic and unpredictable environments, where content popularity is not static and fluctuates based on factors like time, location, and social trends \cite{gao2020design, zong2022cocktail, krishna2021caching, zhang2020deep}. A viral video, for instance, may experience a sudden spike in requests, followed by a rapid decline in demand. Similarly, request rates can change unpredictably, influenced by factors such as user mobility, device connectivity, and network congestion. For instance, during peak hours in a densely populated area, a surge in user activity may dramatically increase the request rate for certain content. On the other hand, during off-peak times, the request rate may drop significantly. Events like sports games or breaking news can also cause rapid shifts in content requests. When it comes to dynamic environments, most existing research primarily focuses on changes in content popularity trends while overlooking the equally important fluctuations in request rates. In real-world scenarios, Users' behavior and usage patterns can shift over time, which directly influences the rate at which files are requested. For instance, during peak hours, when many users are accessing the network at once, demand for content rises sharply. This increased activity results in shorter intervals between file requests, as users require quick access to different resources. In contrast, during off-peak periods, when fewer users are online, the intervals between file requests tend to lengthen. With reduced demand for immediate access, users are less likely to request content frequently, leading to extended gaps between consecutive requests.

Existing caching strategies, particularly those relying on fixed policies, often fail to account for these dynamic shifts, leading to inefficient use of cache resources. Therefore, the ability to learn new policies quickly and efficiently becomes crucial for maintaining high cache performance in such environments.

Most of the existing work on reinforcement learning in dynamic environments assumes that changes occur gradually, giving the agent sufficient time to adapt its policy incrementally. This assumption, however, is not always valid, especially in real-world scenarios where sudden shifts in content popularity or request rates are common. In such cases, relying on gradual adaptation may lead to poor performance. Therefore, detecting when these abrupt changes happen becomes critical to maintaining high system efficiency. By quickly identifying the moment of change, essential steps can be taken to adjust the Reinforcement Learning (RL) policy and prevent the agent from making decisions based on outdated information. Detecting these shifts promptly allows the agent to recalibrate and ensure that caching decisions remain relevant. 

While training caching policies from scratch after every environmental change is one option, it is not practical. Learning a new policy from the ground up each time a shift occurs would result in significant delays, during which the system operates suboptimally. This can lead to increased latency and reduced cache efficiency. For this reason, it is important to converge to the optimal policy faster in the new environment, as this ensures the system can restore its high level of performance without unnecessary delays or inefficiencies.

Building on this, we introduce a caching strategy based on PPO \cite{schulman2017proximal}, a reinforcement learning (RL) technique known for its stability and efficiency. In our approach, we incorporate not only content popularity but also key file attributes such as size, lifetime, and importance to ensure that the cache is used optimally. 

In this work, we propose a mechanism for detecting changes in content popularity and request rates, designed to identify shifts in the environment quickly. This mechanism has two functions. The first function focuses on popularity change detection by leveraging cosine similarity between recent and past request patterns. This method effectively captures shifts in content popularity by comparing how closely current requests align with historical trends. The second function addresses request rate changes using a simple moving average over a defined window of time. By continuously monitoring and averaging request rates, it quickly detects deviations from the norm, signaling a change in the environment. Both functions are designed to detect changes rapidly, allowing the system to adapt swiftly and maintain optimal performance in dynamic scenarios.

To quickly adapt to changing environments, we integrate transfer learning into the PPO algorithm by transferring the learned knowledge in the form of transitions to the new environment. In the new environment, we assign priorities to both the transferred state transitions and the new transitions collected from the current environment. These priorities are determined using a combination of the Temporal difference (TD) error associated with each transition and the difference between the average reward and the instant reward tied to that transition. By assigning higher priority to more relevant transitions, the caching agent is able to focus on learning from the most significant experiences, thus accelerating convergence. This approach is particularly advantageous in dynamic edge caching environments, where frequent changes demand swift adaptation, and leveraging prior knowledge minimizes the time and computational resources required for the agent to adjust to new conditions.

In summary, our approach tackles the limitations of traditional caching strategies by dynamic change detection, and transfer learning. This results in a caching system that is flexible, efficient, and capable of adapting quickly to changing environments, ensuring optimal performance even in highly dynamic scenarios. The contributions of this paper are summarized as follows.

\begin{itemize}

  \item{We introduce a mechanism with two functions specifically designed to rapidly detect changes in the environment. These functions are engineered to enhance the responsiveness of the system to dynamic shifts in content popularity and request rate, ensuring timely and accurate identification of environmental changes.}

  \item{We propose a novel algorithm that combines transfer learning with PPO. This innovative approach facilitates accelerated adaptation to new environments by effectively leveraging previously acquired knowledge, thereby enhancing the algorithm’s ability to quickly adjust and optimize caching strategies in dynamic settings.}

  \item{We conducted simulations where the environment underwent sudden changes, and our change detection approaches demonstrated remarkable speed in identifying these change points. Furthermore, our proposed transfer learning algorithm outperforms several benchmark methods, including learning from scratch, learning from demonstrations \cite{hester2018deep}, and another recently published work \cite{zhou2022learning}. This superiority highlights the effectiveness of our approach in rapidly adapting to new conditions and achieving better performance in dynamic environments.}
  
\end{itemize}

The remainder of the paper is structured as follows: Section \ref{Related Work} provides an overview of related work. Section \ref{System Model} introduces our system model, while Section \ref{The proposed caching algorithm} outlines our proposed caching algorithm. In Section \ref{results}, we discuss our experimental setup and results, and Section \ref{conclusion} concludes the paper. Table \ref{Table of notations} summarises the most important notations used in this paper.

\begin{table}[htbp]
  \centering
  \caption{Table of notations}
  \label{Table of notations}
  \begin{tabular}{lp{0.7\linewidth}}
  \hline
    \textbf{Symbol} & \textbf{Description} \\
    \hline
    $f_r$ & the requested file \\
    $F$ & The total number of file types \\
    $f_f$ & file type $f$ \\
    $\lambda$ & The request rate\\
    $\eta$ & Skewness of popularity distribution \\
    $N$ & length of the sliding window to store the history of the file requests \\
    $\gamma$ & The discount factor \\
    $|\Lambda|$ & The size of the replay buffer before change in the environment\\
    $|\Lambda^{\prime}|$ & The size of the replay buffer after the new environment \\
    $\beta$ & Controls the importance sampling \\
    $d_f$ & The number of requests for file type $f$ over the last N requests \\
    $l_f$ & Lifetime of file type $f$ \\
    $i_f$ & Importance of file type $f$ \\
    $z_f$ & Size of file type $f$ \\
    $h^f(t)$ & Freshness of file type $f$ \\
    $y_f(t)$ & Utility of file type $f$ \\
    $b$ & The files cached in the edge router's cache \\
    $\pi$ & Policy \\
    $\theta$ & Weights of the actor network \\
    $\theta^{\prime}$ & Weights of the critic network \\
    $s$ & Current state \\
    $s^{\prime}$ & Next state \\
    $r(t)$ & The instant reward at time $t$ \\
    $\tau$ & State transtition time \\
    $r_i$ & The instant reward of transition $i$ \\
    $r_{(n)}$ & The immediate reward at the $n^{th}$ step \\
    $p_i$ & The priority of sampling transition $i$ \\
    $P(i)$ & The probability of sampling transition $i$ \\
    $L_P$ & The size of the moving average window for popularity change detection \\
    $L_R$ & The size of the moving average window for request rate change detection \\
    $th_R$ & Threshold for the acceptable variance of the request rate \\
    $th_c$ & Threshold for the acceptable variance of the cosine similarity \\ 
    \hline
  \end{tabular}
\end{table}

\section{Related Work} \label{Related Work}

The literature on caching strategies can broadly be categorized into two main approaches based on how they handle file popularity: those assuming static popularity and those addressing dynamic popularity. In the former, file request patterns are considered fixed over time, with algorithms optimized to perform under the assumption that popular content remains consistent. On the other hand, the latter body of work focuses on environments where file popularity changes over time, necessitating adaptive caching strategies that can respond to shifting user demand. The following sections will review research under these two categories.

\subsection{Caching with Static File Popularity}

Paper \cite{sun2023federateddeep} tackles the challenge of efficient edge caching in multi-tier computing networks, where computing, caching, and networking resources are distributed from the cloud to users through edge servers. The study addresses the high-cost nature of direct cache hits caused by user content requests. To overcome this, the authors propose integrating recommender systems with edge caching in mobile two-tier (edge-cloud) networks to support both direct and soft hits, improving the resource utilization of edge servers. They model the problem as an MDP and introduce a decentralized caching framework using a multi-agent actor-critic algorithm and federated learning. This enables edge servers to independently learn optimal caching strategies while minimizing long-term system costs.

Study \cite{ping_drl_iot} addresses the challenges in IoT networks related to transient data generation, limited energy resources, and quality of service (QoS) requirements by proposing a DRL-based caching strategy. The authors design a hierarchical architecture caching scheme that optimizes both cache hit rate and energy consumption while considering data freshness and the limited lifetime of IoT data. 

The authors in \cite{zhou2023edgecomputation} explore the integration of edge computation offloading and content caching in 6G-enabled Internet of Vehicles (IoV) to tackle the computational and communication demands of emerging vehicular applications. The proposed framework leverages the synergy between computation offloading and caching to reduce latency, improve service quality, and optimize resource usage in highly dynamic IoV environments. 

To address challenges in caching transient IoT data at the edge, researchers in \cite{zhu2018caching} have proposed a DRL-based approach that dynamically manages IoT data caching without prior knowledge of data popularity or user request patterns. The approach considers data freshness as a key metric to balance communication costs and freshness loss, which is essential in IoT applications such as autonomous vehicles, where data must remain timely and accessible

The approach proposed in \cite{ping_drl_iot} improves cache hit rates and reduces energy consumption without requiring prior knowledge of data patterns, tackling issues like data freshness and limited data lifetimes. The scheme uses a hierarchical architecture to capture regional variations in data popularity by deploying edge caching nodes strategically within IoT networks. 

\subsection{Caching with Dynamic File Popularity}

The authors in \cite{zhong2020drlwireless} propose a DRL approach to dynamically adapt caching policies to time-varying content popularity. Using the Wolpertinger architecture, they design actor-critic DRL frameworks for both centralized and decentralized caching scenarios. The centralized policy focuses on maximizing the cache hit rate, while the decentralized approach optimizes both the cache hit rate and transmission delay. 

Study in \cite{DRL_reactive_caching} addresses the issue of managing heavy backhaul traffic generated by user-generated videos, which are characterized by a surge in demand within a short time frame after release. The authors propose a solution that combines content popularity prediction and reactive caching to optimize resource utilization in a three-tier wireless network. To predict popularity in various locations, they employ a recommendation system that translates request probabilities into rating scores. Using these predictions, a deep reinforcement learning-based caching strategy is introduced to make timely caching decisions for newly requested content, aiming to maximize cache efficiency while minimizing storage costs.

Study \cite{yao2021cachedynamic} focuses on optimizing the content placement strategy, aiming to minimize data transmission delay while considering cache storage limitations and data freshness constraints. This problem is formulated as an integer linear programming (ILP) model and later as a MDP, and the authors apply a deep reinforcement learning algorithm to address the dynamic nature of the caching problem.

Study \cite{gao2020reinforcement} addresses the problem of caching in ultra-dense networks with wireless backhaul, focusing on optimizing caching strategies at small base stations (SBSs) with limited storage. The challenge arises from the dynamic and unknown variations in content popularity over time. The authors propose a cooperative caching strategy that leverages RL combined with maximum-distance separable coding to maximize the traffic load SBSs serve, avoiding reliance on the macro base station. The problem is formulated as an MDP, and Q-learning is applied to small-scale systems. They introduce a heuristic approximation of the state-action value function for large-scale systems, significantly reducing the complexity of action selection. 


The authors in \cite{fan2021pa} propose an evolving learning-based content caching policy that adaptively learns the changing content popularity and decides which content to replace when the cache is full. This approach uses a multilayer recurrent neural network (RNN) that evolves its training process from shallow to deeper layers as more requests arrive, offering a dynamic and scalable solution. 

Study \cite{zhang2020deep} addresses the problem of designing an optimal coded caching policy in wireless networks to minimize the total network cost. To tackle the unknown and time-varying content popularity and the continuous, high-dimensional action space, the authors propose a clustering-based long short-term memory (C-LSTM) approach to predict the number of content requests using historical data, leveraging correlations between different files. Based on these predictions, they develop a supervised deep deterministic policy gradient method to learn the caching policy in continuous action space using an actor-critic architecture. 

Authors in \cite{zong2022cocktail} address the issues of dynamic content popularity and heterogeneous caching configurations. The authors propose an ensemble learning approach that adapts to varying caching scenarios by selecting from multiple caching policies, rather than relying on a single dominant policy. They analyze the complementary strengths of Least Frequently Used (LFU) and Least Recently Used (LRU) policies in different situations and develop a caching algorithm that enhances these policies using a deep recurrent neural network (LSTM) for time-series analysis. Additionally, they introduce a deep reinforcement learning agent that dynamically combines base caching policies based on their virtual hit ratios.

Authors in \cite{gao2020design} focus on designing a dynamic probabilistic caching strategy for scenarios where content popularity fluctuates over time, but the average content popularity over a time window can be predicted. The goal is twofold: (i) to converge to optimal caching probabilities when content popularity is stable, and (ii) to adapt to time-varying popularity when it changes. The authors propose a dynamic content replacement method that addresses the limitations of static caching, which cannot adapt, and traditional dynamic policies like LRU, which fail to converge to optimal caching probabilities. The problem is modeled as finding the state transition probability matrix of a Markov chain, and a method is developed to generate and refine this matrix.

Study in \cite{muller2016smart} explores content caching in wireless small-cell networks, where popular files are stored at SBSs to reduce load on the macro cellular network. The authors consider a scenario where popularity varies based on factors such as changing user types. To solve this, they propose a caching algorithm based on contextual multi-armed bandits, allowing the SBS to update its cache dynamically and learn context-dependent popularity profiles over time. 

The authors in \cite{zhou2022learning}
propose a Q-value-based algorithm for joint radio and cache resource allocation in 5G networks, addressing the need for more efficient and flexible resource management.  This approach incorporate knowledge transfer into the double deep Q-network framework, enabling agents to leverage expert knowledge, which accelerates learning and reduces the training burden.

\subsection{Motivation}

While there are research work addressing dynamic content popularity, these works assume a fixed request rate, meaning that the rate at which requests are generated remains constant over time. However, this assumption is not practical for real-world deployments where the request arrival patterns can vary unexpectedly. Ignoring the impact of changing inter-arrival times can result in caching policies that are optimized for specific request patterns but become ineffective when the temporal dynamics shift. This can lead to increased latency and reduced cache hit rate as the system is unable to adapt to the altered demand patterns.

To the best of our knowledge, there is no research in the literature that explicitly addresses the problem of varying request rates in edge caching systems. This is a critical gap because even a caching strategy that adapts to changes in content popularity will struggle when the rate at which requests arrive deviates from what the policy has been trained on. For example, if a policy is tuned for frequent requests but is suddenly faced with infrequent arrivals, it may result in excessive cache replacements or underutilization of resources. Conversely, if the request rate increases significantly, the policy might fail to keep pace with the higher demand, leading to cache saturation and performance degradation.

Therefore, it is essential to consider both dynamic content popularity and varying request rates when designing edge caching policies. A caching strategy that is sensitive to request rate fluctuations can make more informed decisions about when to replace cached content or prefetch new files, ensuring optimal performance under diverse traffic conditions. By capturing both the content dynamics and temporal variations in request patterns, such an approach would be more robust and better suited for real-world edge caching applications, thereby improving cache hit rate, reducing latency, and minimizing the overall network load. This study aims to address this gap and propose a novel solution that effectively handles varying request rates, ensuring adaptive and efficient caching performance in dynamic environments.

Moreover, most of the existing research works do not involve the detection of changes in either content popularity or request rate. Instead, they assume that the exact time points where these changes occur are known beforehand. This assumption simplifies the design of caching strategies, as the models can directly adjust their policies based on predefined change points. However, in practical scenarios, the timing and nature of these changes are typically unknown and can occur unpredictably. Relying on prior knowledge of such changes is unrealistic and limits the applicability of these strategies in real-world environments, where both content popularity and request rates can fluctuate unpredictably. This motivates us to tackle the issue of environmental change detection in our work.

\section{System Model} \label{System Model}

\subsection{System Architecture}

\begin{figure*}
  \centering
  \includegraphics[width=4.5in]{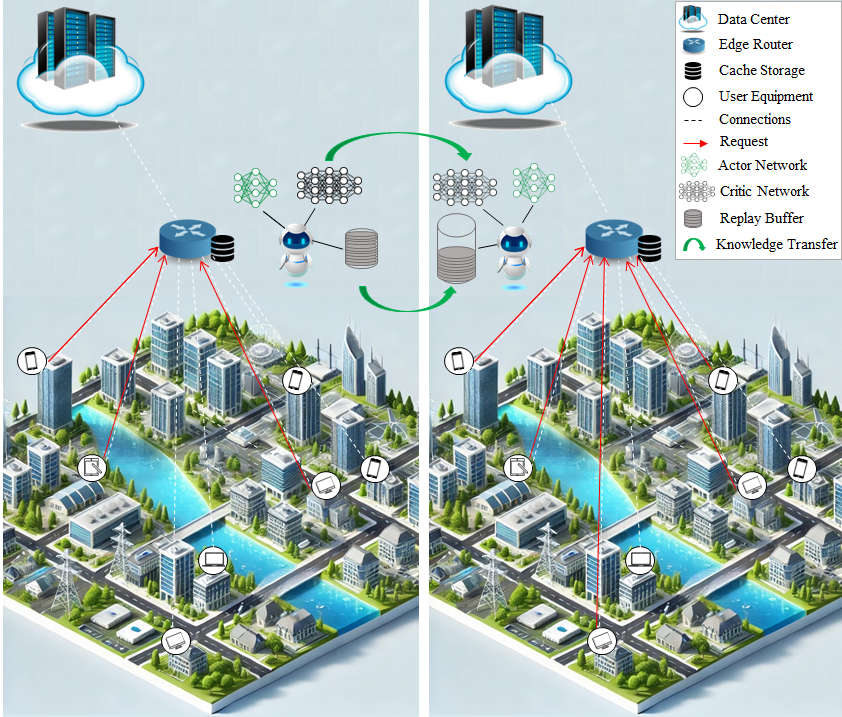}
  \caption{Caching system topology}
  \label{fig_System_Model}
\end{figure*}

We analyze a network environment where cache storage resides within the edge router, which is directly connected to the end users from one side and the data center on another side as shown in Fig.~\ref{fig_System_Model}. The capacity of the edge router is denoted as $M$.

It is assumed that the cloud data center possesses sufficient capacity to store all content \cite{wei2021wireless}. Upon receiving a file request from an end user, a copy is generated and transmitted to the user via the edge router over the Internet. If the requested file $f_r$ is already cached in the edge router, the request is fulfilled directly from its cache without retrieving the file from the data center. From this point forward, the term “cache” refers to the edge router's cache. The end users, represented as devices, request files based on their respective needs and preferences \cite{wei2021wireless}. Let the set of users connected to the edge router be denoted by $\mathcal{U} =\{u_1, u_2, \ldots, u_U\}$. Requests are represented by $\mathcal{G}=\{g_1, g_2, \ldots, g_G\}$, where $g_g$ represents the $g^{th}$ request, irrespective of which user initiates it. Requests are processed in a first-come-first-served manner.

Files, generated from sources such as cameras, sensors, and computers, are stored within the data center. Let $\mathcal{F}=\{f_1, f_2, \ldots, f_F\}$ denote the set of file types, where $f_f$ represents the $f^{th}$ file type. Each file type may exhibit different characteristics, including popularity, lifetime, size, and importance. These features can be further explained using real-life examples:

\begin{itemize}
    \item \textbf{Popularity:} The number of times a file is requested by users \cite{li2016pop}. Popular files, such as trending videos, tend to be requested more frequently than less popular ones.
    
    \item \textbf{Lifetime:} The period during which a file remains valid after leaving the data center \cite{nasehzadeh2020deep}. For example, location-based services often necessitate timely updates, meaning files are assigned a predetermined lifetime upon generation \cite{vural2017caching}.
    
    \item \textbf{Size:} Files vary in size depending on their type and content. For instance, a movie file is generally much larger than a text file.
    
    \item \textbf{Importance:} Represents the value a file holds for users based on its relevance and necessity. Files of critical importance, such as real-time financial data or emergency updates, are more valuable than less critical files like free ebooks or casual videos.
\end{itemize}

A comprehensive caching policy must account for all file characteristics simultaneously. Focusing on only one attribute might neglect other crucial aspects. For example, a highly popular file with a short lifetime, large size, and low importance may be less beneficial than a less popular file with a longer lifetime, smaller size, and higher importance.

This work considers the above four key characteristics for each file. Let $\mathbf{d}(t) = (d_1(t), d_2(t), \dots, d_F(t))$ represent the popularity values for each file type, where $d_f$ corresponds to the popularity of file type $f$. This indicates the number of times file type $f$ is requested within a predefined time interval $N$. Additionally, the lifetime, size, and importance of each file type at time $t$ are denoted as $\mathbf{l}=(l_1, l_2, \ldots, l_F)$, $\mathbf{z}=(z_1, z_2, \ldots, z_F)$, and $\mathbf{i}=(i_1, i_2, \ldots, i_F)$, respectively.

The utility of a cached file, defined as a function of its freshness and importance, determines its overall value. Freshness refers to the age of a file relative to its lifetime at time $t$, expressed as:

\begin{equation*}
h^f(t) = \frac{t - w^f_g}{w^f_l}, \quad 0 \leq h^f(t) \leq 1,
\end{equation*}

\noindent where $t$ denotes the current time, $w^f_l$ is the lifetime, and $w^f_g$ is the file $f$'s generation time.

The utility function for file type $f$ at time $t$ is denoted as $y_f(t) = E(h^f(t), i_f)$, where $y_f(t)$ represents the utility of file type $f$ at time $t$. The function $E$ depends on two essential factors: the file's freshness $h^f(t)$, and the file's importance $i_f$. The function $E$ captures the interaction between these elements in determining the overall utility. Typically, utility $y_f(t)$ increases with the file's importance $i_f$, reflecting its greater value to the system. Moreover, there is an inverse relationship between a file’s freshness and its utility; as freshness $h^f(t)$ decreases, indicating the file is becoming outdated, the utility $y_f(t)$ decreases accordingly. Although the specific definition of a utility function varies across applications and user perspectives, the method proposed here is adaptable and compatible with any utility function definition.

\subsection{System Uncertainties}

Caching involves inherent uncertainties, such as unpredictable request arrivals and the unknown impact of future requests on the cache. These uncertainties influence the edge router's decision-making process. Below, we describe the primary uncertainties and our corresponding assumptions.

\subsubsection{Change in Popularity}

In dynamic environments such as edge caching systems, the popularity of cached content can fluctuate over time. This change in popularity refers to the variation in how often users request certain files or data. Various factors, such as user preferences, temporal trends, and external events typically influence the popularity distribution. As the popularity changes, files that were previously requested frequently may become less popular, while others gain prominence. Managing this uncertainty is crucial to maintaining optimal cache performance, as the caching policy must continuously adapt to the shifting popularity of content to reduce latency and minimize cache misses.

\subsubsection{Change in Request Rate}

Another major source of uncertainty in our system is the variability in the request rate, which refers to the number of file requests per unit of time. The request rate can fluctuate based on user behavior, network conditions, or time of day, and these fluctuations can impact the performance of the caching system. A sudden spike in requests, for example, may overload the system, leading to increased latency. On the other hand, a reduction in the request rate may result in under-utilized cache resources. We model request arrivals using a Poisson process,
a common model for characterizing user request patterns \cite{gomaa2013estimating}.
The parameter \(\lambda\) represents the request rate. We assume that the edge router has no prior knowledge of the
Poisson process or its parameters.

\section{The proposed caching algorithm} \label{The proposed caching algorithm}

\subsection{SMDP Formalism}

The responsibility of the edge router involves deciding which files should be cached, taking into account factors such as user request frequencies, file properties, and cache memory constraints. This decision-making procedure is inherently sequential, making MDPs a suitable modeling approach, as outlined in \cite{puterman2014markov}.

An MDP is defined by five key elements: state, action, system dynamics, reward, and policy. In this context, a state represents the condition of the system at a given time instance. The agent selects an action following a defined policy, transitions into a new state, and receives a reward based on the action's quality. This process repeats either indefinitely (infinite horizon) or until a termination point is reached (finite horizon) \cite{sutton2018reinforcement}. In the caching problem, the infinite-horizon approach is applicable since no predefined termination state exists. As requests arrive continuously, the problem requires a continuous-time extension of the MDP. Therefore, we employ the SMDP framework, which can handle variable transition times \cite{puterman2014markov}. We follow our previous work for SMDP formulation presented in \cite{niknia2023edge} Below, we briefly explain the components of out SMDP formulation and its components.

An SMDP is characterized by the tuple $(\mathcal{S}, \mathcal{A}, J, R, \pi)$, where $\mathcal{S}$ denotes the state space, $\mathcal{A}$ represents the action space, $J$ signifies transition durations, $R$ is the reward function, and $\pi$ is the policy.

\subsubsection{System States}

At time $t$, the system state is given by $s_t \in \mathcal{S}$ and includes:

\begin{equation*}
\begin{aligned}
s(t) = \{ & Mem(t), \; \mathbf{b}(t), \;  \mathbf{y}(t), \;
\mathbf{d}(t), \; \mathbf{i}(t), \; \mathbf{l}(t), \; \mathbf{z}(t) \},
\end{aligned}
\end{equation*}

\noindent where $Mem(t)$ represents the available cache memory in the edge router. Their values are defined as follows:

\begin{equation*}
Mem(t) = \frac{M-\sum_{f=1}^F{\mathbf{b}(t)\cdot \mathbf{z}(t)}}{M},
\end{equation*}

\noindent where $\mathbf{b}(t)$ is a binary vector, each element indicating whether a specific file is cached in the edge router. This vector is defined as $\mathbf{b}(t) \in B = \{0, 1\}^F$, with 1 representing a cached file and 0 representing an uncached file. The vector $\mathbf{y}(t)$ tracks the utility of cached files in the edge router. IF $y_f(t) = 0$, the file $f$ is not cached. Additionally, $\mathbf{i}(t)$, $\mathbf{l}(t)$, and $\mathbf{z}(t)$ represent the importance, lifetime, and size of each file, respectively.

\subsubsection{Actions}

At any time $t$, the agent selects one of the two possible actions: $a(t) = 1$ (cache the file in the cache), or $a(t) = 0$ (do not cache the file). If a cache is full, the file with the lowest utility is evicted to create space for new files. Multiple files can be removed if further space is required.

\subsubsection{System Dynamics}

In this SMDP model, with stochastic task arrivals, the system's evolution depends on both state transition probabilities $(P_{ss^\prime})$ and the associated transition times $(\tau)$. The inclusion of time alongside state transitions increases the complexity of decision-making, compared to traditional MDPs. If the system dynamics and the reward function were fully known, Bellman equations could be applied to determine the optimal policy. However, as this information is typically unknown in practical applications, reinforcement learning is employed to iteratively learn the system dynamics and optimize the policy through experience. The specifics of the RL algorithm used will be described in subsequent sections.

\subsubsection{Reward}

The immediate reward is expressed as:

\begin{equation}
\begin{aligned}
r(t) = & \; w_1 \left[ \left( \mathbf{b}(t) \cdot \mathbf{d}(t) \right) \left( \mathbf{b}(t) \cdot \mathbf{y}(t) \right)^T  \; \right] \\
& \; - w_2 \left[ Mem(t) \right],
\end{aligned}
\label{reward function}
\end{equation}

\noindent where the first term represents the weighted utility of cached files, and the second term reflects the unused portion of cache memory. The long-term objective is to maximize the average cumulative utility of cached files while minimizing the average unused cache capacity.

\subsubsection{Policy}

The policy $\pi$ dictates the optimal action in each state to achieve the long-term objective.

\subsection{Proximal Policy Optimization} \label{RL algorithm}

PPO is a widely adopted reinforcement learning method, appreciated for its robustness and simplicity. The primary innovation of PPO lies in mitigating the issues of policy updates through a constrained objective, thereby ensuring more controlled updates during training.

PPO works by optimizing the expected reward while constraining the difference between the updated and previous policies. The surrogate objective used in PPO is formulated as follows:

\begin{equation}
\begin{aligned}
L^{\text{PPO}}(\theta) = \min \Bigg( & \frac{\pi_\theta(a(t) \mid s(t))}{\pi_{\theta_\text{old}}(a(t) \mid s(t))} \hat{A}(t), \\
& \text{clip} \left( \frac{\pi_\theta(a(t) \mid s(t))}{\pi_{\theta_\text{old}}(a(t) \mid s(t))}, 1 - \epsilon, 1 + \epsilon \right) \hat{A}(t) \Bigg) ,
\label{LPPO}
\end{aligned}
\end{equation}

\noindent where $\pi_\theta(a(t) \mid s(t))$ represents the probability of selecting action $a(t)$ in state $s(t)$ under the policy parameterized by $\theta$, $\pi_{\theta_\text{old}}(a(t) \mid s(t))$ is the probability under the prior policy, $\hat{A}(t)$ is the advantage function, and $\epsilon$ is a clipping hyperparameter.

The policy is updated by maximizing this clipped surrogate objective using stochastic gradient ascent. This clipping mechanism limits the extent to which the new policy differs from the old one, maintaining a balance between policy exploration and exploitation.

The advantage function $\hat{A}(t)$ indicates the relative benefit of an action in comparison to a baseline:

\begin{equation}
\hat{A}(t) = r(t) + \gamma^\tau V_{\theta ^ \prime}(s^\prime) - V_{\theta ^ \prime}(s),
\end{equation}

\noindent where $V_{\theta ^ \prime}(s)$ and $V_{\theta ^ \prime}(s^\prime)$ are the values of the current and next states, respectively. In standard MDPs, $\tau$ is set to 1 due to uniform state transition times, simplifying the equation with $\gamma^1$. In contrast, SMDPs involve variable transition times, so $\tau$ accounts for the transition time between states. In this case, $\gamma$ is raised to the power of $\tau$, adjusting for the varied time intervals and ensuring appropriate discounting of future rewards over irregular time steps \cite{SMDP, niknia2022thermal}.

In addition to the policy update, the value function is trained using a separate loss term, defined as the squared difference between the predicted state value $V_{\theta ^ \prime}(s)$ and the target return $R(t)$:

\begin{equation}
\label{original value loss}
L^{\text{value}}(\theta ^ \prime) = \left( V_{\theta ^ \prime}(s) - R(t) \right)^2 .
\end{equation}

This loss ensures that the value network provides accurate estimates for state values during policy training.

The target return $R(t)$ is given by the one-step return:

\begin{equation}
R(t) = r(t) + \gamma^\tau V_{\theta ^ \prime}(s^\prime).
\end{equation}

This formulation ensures the value network adjusts by minimizing the error between the estimated value and the return, derived from both the immediate reward and the discounted future value.

\subsection{Change Detection}

Until here, we introduced our proposed caching algorithm that can learn the optimal caching policy for a given environment, considering the request rate and content popularity distribution. As discussed earlier, this learned policy may no longer remain optimal if the request rate or content popularity changes. Such changes can occur at unpredictable times, which makes it crucial to detect these shifts as quickly as possible to ensure the caching policy remains effective.

We propose our change detection mechanism for detecting changes in request rate and content popularity. One function of this mechanism focuses on monitoring fluctuations in the request rate to identify when the current pattern deviates significantly from the learned model, indicating a change. The other function targets changes in content popularity by tracking variations in the distribution of content requests. By implementing these detection functions, our approach can dynamically identify and respond to changes in the environment, allowing the caching strategy to adapt and maintain optimal performance.

\subsubsection{Request Rate Change Detection}

In our research work, we implement a request rate change detection function by monitoring the interarrival times of user requests. This is achieved by calculating a moving average over a specified window of length $L_R$, denoted as $W_{R}$. Let the interarrival times of user requests be represented by $\{\tau_1, \tau_2, \ldots, \tau_c, \ldots, \tau_C\}$, where $\tau_c$ is the time difference between the $c$-th and $(c-1)$-th request arrival. The moving average $W_{R}$ over a window of size $L_R$ is defined as:

\begin{equation}
W_{R} = \frac{1}{L_R} \sum_{j=c-L_R+1}^{c} \tau_j.
\end{equation}

This moving average $W_{R}$ provides a smoothed estimate of the recent interarrival times, capturing the overall trend in request frequency. The goal is to detect any significant deviation in $W_{R}$ from the average interarrival times of requests in the previous environment denoted by $1/\lambda$, which represents the expected interarrival time under normal conditions. To achieve this, we define a threshold for acceptable variance, denoted as $th_{R}$. The condition for detecting a significant change in request rate is given by:

\begin{equation}
|W_{R} - 1/\lambda| > th_{R}.
\end{equation}

If the absolute difference between $W_{R}$ and $1/\lambda$ exceeds $th_{R}$, it indicates a notable change in the request arrival pattern, implying that the request rate has varied significantly from the expected behavior. This condition can occur in scenarios where, for example, a sudden increase in user demand causes a spike in the request rate, or a decrease in user engagement leads to fewer incoming requests over time.

\subsubsection{Content Popularity Change Detection}

We propose a method for detecting changes in content popularity by keeping track of request patterns over recent time intervals. This is accomplished using a history of requests spanning the most recent $L_P$ requests. We define a vector $W_P = \{W_1, W_2, W_3, \ldots, W_f, \ldots, W_F\}$, where $W_f$ is the average number of times that file $f$ is requested within the interval spanning from trials $n - L_P$ to $n$. Formally, $W_f$ can be represented as:

\begin{equation}
W_f = \frac{1}{L_P} \sum_{i=n-L_P}^{n} I_f(i),
\end{equation}

\noindent where $I_f(i) = 1$ if file $f$ is requested at trial $i$, and $I_f(i) = 0$ otherwise. Thus, $W_f$ captures the relative popularity of file $f$ over the most recent $L_P$ trials.

Similarly, we define $W_P^{\text{old}}$ as the vector of average request counts for files over the previous $L_P$-length interval, spanning from trials $n - L_P - 1$ to $n - 1$:

\begin{equation}
W_f^{\text{old}} = \frac{1}{L_P} \sum_{i=n-L_P-1}^{n-1} I_f(i).
\end{equation}

\textbf{Cosine Similarity for Popularity Change Detection:} To detect changes in popularity, we compute the cosine similarity between $W_P$ and $W_P^{\text{old}}$. Cosine similarity measures the angle between two vectors in the multi-dimensional space, defined as:

\begin{equation}
\text{Cosine Similarity}(W_P, W_P^{\text{old}}) = \frac{W_P \cdot W_P^{\text{old}}}{\|W_P\| \cdot \|W_P^{\text{old}}\|},
\end{equation}

where:

\begin{itemize}
    \item $W_P \cdot W_P^{\text{old}} = \sum_{f=1}^{F} W_f W_f^{\text{old}}$ is the dot product of the two vectors.
    \item $\|W_P\| = \sqrt{\sum_{f=1}^{F} W_f^2}$ and $\|W_P^{\text{old}}\| = \sqrt{\sum_{f=1}^{F} \left(W_f^{\text{old}}\right)^2}$ are the magnitudes (or norms) of $W_P$ and $W_P^{\text{old}}$, respectively.
\end{itemize}

The cosine similarity value ranges from -1 to 1, where:

\begin{itemize}
    \item 1 indicates that the two vectors are identical in direction (no change in popularity).
    \item 0 indicates that the vectors are orthogonal (completely different popularity patterns).
    \item -1 indicates that the vectors are in opposite directions.
\end{itemize}

\textbf{Rationale for Using Cosine Similarity:} Cosine similarity is well-suited for popularity change detection because it captures changes in the relative proportions of requests for different files, irrespective of their absolute magnitudes. This makes it effective in distinguishing shifts in content demand patterns, even when the total request volume remains constant. For instance, if previously popular files become less requested and other files increase in popularity, the cosine similarity will capture this shift, while other measures like Euclidean distance might not be as sensitive to these relative changes.

\textbf{Change Detection Mechanism:} Once the cosine similarity value is computed, we maintain a moving window of these similarity values, denoted as $W_C$. Let $W_C = \{ C_1, C_2, \ldots, C_m \}$, where $C_m$ is the cosine similarity value calculated at the $m$-th interval. If the average of the values in $W_C$ falls below a predefined threshold $th_C$, it indicates that the similarity between $W_P$ and $W_P^{\text{old}}$ has decreased significantly over time, suggesting a change in the popularity distribution:

\begin{equation}
\frac{1}{M} \sum_{m=1}^{M} C_m < th_C,
\end{equation}

\noindent where $M$ is the length of the moving window. If this condition is met, we conclude that a significant change in content popularity has occurred, and the caching strategy should be updated accordingly.

This approach is efficient and responsive because it does not require prior knowledge of when or how the popularity will change, making it robust for real-world scenarios where such information is typically unavailable.

\subsection{Transfer Learning with Demonstrations}

When a change in the environment is detected, it is crucial to learn a new policy that can effectively adapt to the new dynamics. To facilitate this transition, we propose a PPO-based transfer learning mechanism that utilizes previously collected transitions from the old environment as demonstration data in the training process of the new policy. 

\subsubsection{Replay buffer with demonstration data}

The demonstration transitions are stored in the first half of the replay buffer in the new environment ($\Lambda ^ \prime$), while the agent begins exploring the new environment with a high initial exploration rate (\(\epsilon = 1\)). During this exploration, it generates new transitions to fill the second half of the buffer. The full replay buffer provides a mix of old and new experiences. Mathematically, the buffer capacity is defined as \( |\Lambda ^ \prime| \), where \( |\Lambda ^ \prime| = 2 |\Lambda| \). $|\Lambda|$ represents the size of the replay buffer before the change in the environment. As the agent continues to interact with the new environment, the oldest demonstration data is gradually overwritten by new transitions.

Overwriting the demonstration data with new transitions is beneficial for several reasons. Initially, demonstration data from the previous environment serves as a valuable starting point, stabilizing the training process by providing the agent with relevant knowledge. However, as the agent continues to explore the new environment, these old transitions become less useful because they do not accurately represent the current dynamics and may lead to suboptimal learning. By gradually replacing demonstration data with new transitions, the agent ensures that its learning focuses on the most up-to-date information, allowing it to adapt more effectively to the changes in the environment. This process strikes a balance between leveraging prior knowledge for a faster start and transitioning fully to the new environment, ultimately improving the overall policy performance and convergence speed.

To further optimize the adaptation process, we propose reinitializing the actor neural network with random values while importing the critic neural network from the previous environment and fine-tuning it. The rationale behind this approach is to promote exploration by allowing the actor to discover new actions without being biased by the outdated policy while leveraging the previously learned value function to guide learning and accelerate convergence in the new environment. This strategy balances the need for policy adaptation with value stability, ultimately speeding up the learning process.

This approach balances the exploration-exploitation trade-off by initializing the actor network for adaptability and preserving the critic's value estimations for stability. As the replay buffer updates dynamically, the algorithm effectively transitions from demonstration-driven learning to fully online learning in the new environment, thereby improving convergence speed and robustness against abrupt changes.

\subsubsection{Prioritized batch sampling}

We use the following steps to compute priority of sampling of each transition stored in the replay buffer:

\begin{itemize}
    \item \textbf{Calculate Priorities}: The priority $p_i$ of a transition is obtained as:

    \begin{equation}
    p_i=\left(\operatorname{avg}_r - r_i\right)+|TDE_i|+\varsigma,
    \label{priorities ours}
    \end{equation}

    \noindent where $r_i$ is the instant reward of transition $i$ and it is obtained by Eq. (\ref{reward function}). we calculate $\operatorname{avg}_r$ as follows:

    \begin{equation*}
    \operatorname{avg}_r = \frac {\sum_{i=0}^{t} r_i}{t}.
    \end{equation*}

    $TDE_i$ is the temporal difference error that transition $i$ could make the last time it was sampled and it is obtained by:
    
    \begin{equation*}
    TDE_i = V_{\theta^\prime}(s) - R(t).
    \end{equation*}
    
    Moreover, $\varsigma$, is a small positive constant that ensures all transitions have a chance of being sampled.

   \item \textbf{Update Probabilities}: The probability of transition $i$ being sampled is:

    \begin{equation*}
    P(i)=\frac{p_i^{\alpha}}{\sum_{e \in \Lambda} p_e^\alpha},
    \end{equation*}
    
    \noindent where the value of $\alpha$ determines the degree of prioritization, with $\alpha = 0$ corresponding to the uniform sampling case, i.e., transition $i$ is sampled randomly.
    To adapt to environment change, adjustments are made by applying importance sampling weights, as illustrated below:
    
    \begin{equation*}
    \omega_i=\left(\frac{1}{\left|\Lambda\right|} \cdot \frac{1}{P(i)}\right)^\beta,
    \end{equation*}
    
    \noindent where $\beta$ controls the degree of importance sampling. When $\beta$ is 0, there is no importance sampling, whereas when $\beta$ is 1, full importance sampling is employed.
    $\omega_i$ is multiplied by the loss function to control the impact of each transition on updating the neural network \cite{hester2018deep}.

\end{itemize}

Overwriting demonstration data and using the proposed prioritized batch sampling help with faster convergence to the optimal policy in the new environment. The rationale behind designing this prioritized batch sampling lies in its ability to emphasize more informative experiences—specifically, experiences that provide higher learning value for the current policy. By selectively sampling transitions expected to contribute most to the policy update, prioritized batch sampling guides the learning process more effectively than uniform sampling. 

In dynamic environments where state-action distributions shift, certain transitions become significantly more relevant to the new optimal policy. Prioritized sampling identifies these valuable transitions, ensuring they are more frequently revisited. This targeted focus allows the agent to adapt more rapidly to environmental shifts, effectively forgetting outdated information while reinforcing knowledge that aligns with the new environment’s dynamics. Consequently, this mechanism accelerates the convergence toward the optimal policy, as it reduces the time and computational resources spent on learning from less relevant past experiences.

\subsubsection{The actor loss function}
We modify the loss function of the actor network in Eq. (\ref{LPPO}) as follows:

\begin{equation}
\begin{aligned}
L^{\text{actor}}(\theta) = L^{\text{PPO}}(\theta) + \lambda_1 J_E(\pi),
\end{aligned}
\end{equation}

\noindent where $J_E(\pi)$ is a large margin loss designed to encourage the policy \( \pi \) to favor actions that align with expert demonstrations, thereby enhancing the learning process in scenarios where guidance from an expert is available. $J_E(\pi)$ is defined as:

\begin{equation}
J_E(\pi) = \max_{a \in A} \left[ \pi_{\theta}(a | s) + l(a_E, a) \right] - \pi_{\theta}(a_E | s),
\end{equation}

\noindent where \( \pi_{\theta}(a | s) \) denotes the probability of selecting action \( a \) given state \( s \) under the current policy parameterized by \( \theta \), and \( l(a_E, a) \) is a margin penalty designed to encourage the agent to favor actions aligned with expert demonstrations. The margin penalty is defined such that it assigns a constant value for actions other than the expert action \( a_E \), effectively boosting the adjusted probability for non-expert actions and incentivizing the selection of \( a_E \). The contribution of \( J_E(\pi) \) is set to zero for self-generated transitions, allowing the agent to explore its environment without being overly constrained by expert behavior. This design promotes a balance between imitation of expert behavior and the flexibility to discover potentially superior strategies through exploration, enhancing the overall robustness and adaptability of the learning process.

\subsubsection{The critic loss function}

We modify the loss function of the critic network in Eq. (\ref{original value loss}) as follows:

\begin{equation}
L^{\text{value}}(\theta ^ \prime) = \mathcal{L}_{\text{1-step}} + \lambda_2 \mathcal{L}_{\text{n-step}} + \lambda_3 \mathcal{L}_{\text{reg}},
\end{equation}

\noindent
This updated loss function incorporates different components, including 1-step and n-step Temporal Difference (TD) losses, and an L2 norm regularization term. The parameters $\lambda_2$ and $\lambda_3$ control the relative contributions of each term to the overall loss function. By integrating these elements, the critic network can effectively learn from both immediate and long-term rewards, while avoiding overfitting.

The 1-step TD error, denoted as $\mathcal{L}_{\text{1-step}}$, is defined as:

\begin{equation*}
\mathcal{L}_{\text{1-step}} = \left( V_{\theta ^ \prime}(s) - R(t) \right)^2. 
\end{equation*}

The n-step TD loss, represented as $\mathcal{L}_{\text{n-step}}$, is expressed as:

\begin{multline}
\mathcal{L}_{\text{n-step}} = \left(R(t)_{(n)} + \gamma^{\tau_{(n)}} V_{\theta ^ \prime} (s^\prime_{(n)}) 
- V_{\theta ^ \prime}(s) \right) ^2,
\end{multline}

\noindent where $V_{\theta ^ \prime} (s^\prime_{(n)})$ refers to the state encountered after $n$ steps, and $\tau_{(n)}$ is the cumulative transition time from $s$ to $s^\prime_{(n)}$. The cumulative discounted reward $R(t)_{(n)}$ is calculated as:

\begin{equation*}    
\begin{gathered}
R(t)_{(n)}=\gamma^{\tau_{(1)}} r_{(1)}+\gamma^{\tau_{(2)}} r_{(2)}+\gamma^{\tau_{(3)}} r_{(3)}+\ldots+\gamma^{\tau_{(n)}} r_{(n)} \\
\tau_{(n)}=t_1+t_2+t_3+\ldots+t_n,
\end{gathered}
\end{equation*}

\noindent where $t_n$ represents the transition time from state $s^\prime_{(n - 1)}$ to $s^\prime_{(n)}$. $r_{(n)}$ is the immediate reward at the $n^{\text{th}}$ step. Incorporating n-step returns ensures that the values of subsequent states are propagated back to preceding states, enhancing the initial training process.

$\mathcal{L}_{\text{reg}}$ is an L2 regularization term designed to reduce overfitting by penalizing large weights.

\section{Experimental Setup and Results} \label{results}

This section starts by outlining the experimental setup and detailing the tools employed for implementing our proposed caching approach called Transfer Learning with Prioritization (TLP). Following that, we present two state-of-the-art transfer learning algorithms, which will act as baseline models for performance evaluation.

We implemented our system model and trained the DRL agent using Python 3. To facilitate the creation and management of neural networks, we utilized the TensorFlow framework as mentioned in \cite{developers2022tensorflow}.

\subsection{Configuration and Parameters}

We exploit two separate neural networks for the actor and critic components. The network weights were initialized uniformly between -0.1 and 0.1, and the biases were set to a constant value of 0.1. The ReLU function was chosen as the activation function. Unless otherwise stated, the setting of the parameters is provided in Table \ref{table-results}.

\begin{table}[htbp]
  \centering
  \caption{parameter settings}
  \label{table-results}
  \begin{tabular}{lp{0.5\linewidth}}
    \hline
    \textbf{Notation}& \textbf{value}\\
    \hline
    $F$ & 50 \\
    $\gamma$ & 0.99 \\
    $M$ & 5000 \\
    $\beta$& 0.6 linearly increased to 1 \\
    $\alpha$ & 0.4 \\
    Batch size & 64 \\
    $|\Lambda|$ & 2000 \\
    $|\Lambda ^ \prime|$ & 4000 \\
    $l$ for each file type & Randomly generated from [10, 30] \\
    $i$  for each file type & Randomly generated from [0.1, 0.9] \\
    $z$  for each file type & Randomly generated from [100, 1000] \\
    $\lambda$ & [0.2, 0.3] \\
    $\eta$ & [0, 1] \\
    $L_R$ & 10 \\
    $L_P$ & 50 \\
    $th_R$ & 0.05 \\
    $th_c$ & 0.3 \\ 
    \hline
  \end{tabular}
\end{table}

The file request probabilities are modeled using a Zipf distribution, governed by the parameter $\eta$ within the range $0 < \eta \leq 1$. According to this distribution, the probability of requesting the $f^{th}$ file is expressed as $\mathfrak{p}_f = \frac{1}{\sigma f^\eta}$ \cite{gomaa2013estimating}, where $\sigma$ is calculated as follows:

\begin{equation}
\sigma = \sum_{f=1}^F {\frac{1}{f^\eta}},
\end{equation}

\noindent The parameter $\eta$ dictates the skewness of the Zipf distribution. As $\eta$ nears 1, the probability of requesting the most frequently accessed file becomes much higher relative to other files. On the other hand, when $\eta$ is closer to 0, the distribution of file popularity tends to be more uniform. The utility function is designed to grow linearly based on a file's significance and increases exponentially with its freshness.

\subsection{Baselines}

To evaluate our proposed approach, we establish baselines that encompass both change detection and transfer learning methods within reinforcement learning environments.

\textbf{Change Detection Baselines:}
\begin{itemize}
    \item \textbf{Kullback-Leibler Divergence (KL Divergence)}: This statistical measure the difference between two probability distributions. It computes the divergence by assessing how one distribution diverges from a second, reference distribution. Specifically, KL Divergence is defined as \( D_{KL}(P || Q) = \sum_{i} P(i) \log \left( \frac{P(i)}{Q(i)} \right) \), where \( P \) and \( Q \) represent the two distributions in question. A higher KL divergence value signifies a more substantial discrepancy between the two distributions.
\end{itemize}

\textbf{Transfer Learning Baselines:}
\begin{itemize}
    \item \textbf{Learning from Scratch (LFS)}: Upon detecting a change in the environment, the agent resets its replay buffer and reinitializes its neural network. It then begins the process of learning the optimal policy from the ground up, without leveraging any prior knowledge from previous policies.
    
    \item \textbf{Deep Q-learning from Demonstrations (DQfD)} \cite{hester2018deep}: DQfD imports its policy from the previous environment and subsequently fine-tunes it within the new context. This method assigns priorities to experiences based on the temporal difference error and retains demonstration data from earlier environments without overwriting it. DQfD is built upon the framework of Double Deep Q-learning.
    
    \item \textbf{Q-value based Deep Transfer Reinforcement Learning (QDTRL)} \cite{zhou2022learning}: Like DQfD, QDTRL is grounded in the principles of DDQL, but it adopts a different approach to transfer learning. This method does not import demonstration data from prior environments and instead samples experience batches randomly without any prioritization. QDTRL incorporates the Q-values derived from the previous environment's policy as an additional term in its loss function, effectively using these Q-values as prior knowledge to guide the learning process.
\end{itemize}

\subsection{Evaluation Criteria}

The performance of our proposed approach is assessed using the following evaluation criteria:

\begin{itemize}

    \item \textbf{Detected Change Point:}  
    One of the key goals is for the agent to detect the change point in the environment as promptly as possible. Early detection allows the agent to transition swiftly into the transfer learning phase, thereby enabling it to adapt its policy to the new environment more efficiently. The effectiveness of different change detection methods is measured by how quickly they identify changes in the environment. The faster the change point is detected, the less time is wasted on the outdated policy, resulting in improved adaptability and overall system performance. The algorithm that minimizes this detection delay is considered the most effective for change detection.

    \item \textbf{Rate of Convergence:}  
    Once the agent detects a change in the environment and initiates the transfer learning process, the rate at which it converges to the optimal policy becomes crucial. The convergence rate refers to how rapidly the agent stabilizes at its peak average reward in the new environment. A high rate of convergence is indicative of the algorithm's efficiency in learning and adapting to new conditions. 

    \item \textbf{Average reward:} For each training trial, the average of all instantaneous rewards accumulated up to that point is computed. The primary objective of an RL agent operating in an infinite horizon is to maximize this average reward over time. Since a higher reward corresponds to a higher cache hit rate, an algorithm that achieves a higher long-term average reward demonstrates superior performance.

\end{itemize}

\subsection{Experimental Findings}

In this section, we present our experimental results and provide a comprehensive comparison with the baseline methods. We first analyze the performance of the change detection mechanism, focusing on their ability to promptly identify shifts in the environment. This evaluation emphasizes detection speed and accuracy, benchmarking our approach against traditional methods to highlight its effectiveness. Next, we shift our attention to the transfer learning algorithms, examining their ability to quickly adapt and stabilize in the new environment once a change has been detected. We compare their convergence rates and learning efficiency against other benchmark algorithms, demonstrating how our proposed method outperforms existing solutions in terms of adaptability and overall performance.

\paragraph{Performance Evaluation of Change Detection Mechanism}

In this subsection, we present the results of our proposed change detection mechanism designed to detect changes in content popularity. The change in file popularity occurs at trial 1000. Tables \ref{cosine} show the specific trials at which our change detection mechanism detected these changes under different experimental scenarios. We varied the number of file types (50, 20, and 10) and considered three distinct cases: 1) Exchanging the popularity of the most and the least popular files, 2) Exchanging the popularity of two popular files with two less popular files, and 3) Exchanging the popularity of five popular files with five less popular files.

Table \ref{cosine} showcases the performance of our change detection mechanism using cosine similarity under various scenarios. The results indicate that setting the threshold ($th_c$) to 0.5 is problematic, as it leads to premature detections well before the actual change point at trial 1000. For example, in multiple cases, the algorithm incorrectly identifies a change as early as trial 51, highlighting the need for a lower threshold value.

Reducing the threshold to 0.4 enhances detection accuracy for environments with 10 and 20 file types, allowing the algorithm to accurately detect changes within 23-33 trials after the true change point. However, for environments with 50 file types, this threshold still triggers false alarms. In contrast, a threshold of 0.3 delivers the best performance across all file types and scenarios, achieving reliable detection without introducing false positives, making it the most suitable option for identifying shifts in popularity trends.

\begin{table}[h] 
\centering
\caption{Change Detection Results for Cosine Similarity with Different Thresholds}
\begin{tabular}{|c|c|c|c|c|}
\hline
\textbf{Change Scenario} & \textbf{$\mathbf{th_c}$} & \textbf{$\mathbf{F = 50}$} & \textbf{$\mathbf{F = 20}$} & \textbf{$\mathbf{F = 10}$} \\ \hline

\noalign{\hrule height 1.2pt} 

2 Popular and & 0.2 & 1056 & 1060 & 1063 \\ \cline{2-5} 
2 Less Popular & 0.3 & 1044 & 1041 & 1041 \\ \cline{2-5} 
Files Exchanged & 0.4 & 701 & 1024 & 1033 \\ \cline{2-5} 
 & 0.5 & 51 & 51 & 51 \\ \hline

\noalign{\hrule height 1.2pt} 

Most and Least & 0.2 & 1049 & 1046 & 1052 \\ \cline{2-5} 
Popular Files & 0.3 & 1042 & 1035 & 1040 \\ \cline{2-5} 
Exchanged & 0.4 & 701 & 1023 & 1032 \\ \cline{2-5} 
 & 0.5 & 51 & 51 & 51 \\ \hline

\noalign{\hrule height 1.2pt} 

5 Popular and & 0.2 & 1045 & 1041 & 1047 \\ \cline{2-5} 
5 Less Popular & 0.3 & 1040 & 1031 & 1037 \\ \cline{2-5} 
Files Exchanged & 0.4 & 701 & 1021 & 1029 \\ \cline{2-5} 
 & 0.5 & 51 & 51 & 51 \\ \hline

\end{tabular}
\label{cosine}
\end{table}

Table \ref{kl} illustrates the results of our change detection algorithm utilizing KL divergence. Selecting an appropriate threshold for KL divergence is inherently complex, as it is largely contingent upon the specific differences between the two environments. For example, when the change involves the popularity of five popular files and five less popular files, a threshold of 10 performs satisfactorily across all file types. Conversely, in scenarios where the change consists of the popularity of the most and least popular files, a threshold of 10 is ineffective. In such cases, the algorithm may fail to detect the change entirely or only identify it after trial 473, which is too late. In practical, real-world situations, there is often a lack of prior knowledge regarding the exact nature of the differences between environments. This uncertainty makes KL divergence susceptible to inaccuracies, thereby rendering it a less reliable choice for change detection in this context.

\begin{table}[h] 
\centering
\caption{Change Detection Results for KL Divergence with Different Thresholds}
\begin{tabular}{|c|c|c|c|c|}
\hline
\textbf{Change Scenario} & \textbf{$\mathbf{Threshold}$} & \textbf{$\mathbf{F = 50}$} & \textbf{$\mathbf{F = 20}$} & \textbf{$\mathbf{F = 10}$} \\ \hline

\noalign{\hrule height 1.2pt} 

2 Popular and & 2 & 51 & 51 & 736 \\ \cline{2-5} 
2 Less Popular & 4 & 73 & 1047 & 1055 \\ \cline{2-5} 
Files Exchanged & 8 & 1061 & 1063 & 1055 \\ \cline{2-5} 
 & 10 & 1473 & -- & -- \\ \hline

\noalign{\hrule height 1.2pt} 

Most and Least & 2 & 51 & 51 & 736 \\ \cline{2-5} 
Popular Files & 4 & 73 & 1047 & 1055 \\ \cline{2-5} 
Exchanged & 8 & 1155 & 1065 & 1055 \\ \cline{2-5} 
 & 10 & 1048 & 1126 & 1419 \\ \hline

\noalign{\hrule height 1.2pt} 

5 Popular and & 2 & 51 & 51 & 736 \\ \cline{2-5} 
5 Less Popular & 4 & 73 & 1027 & 1055 \\ \cline{2-5} 
Files Exchanged & 8 & 1049 & 1072 & 1287 \\ \cline{2-5} 
 & 10 & 1048 & 1050 & 1055 \\ \hline

\end{tabular}
\label{kl}
\end{table}

Based on the results in Tables \ref{cosine} and \ref{kl}, it is evident that cosine similarity provides a more straightforward thresholding approach compared to KL Divergence. While both methods can effectively detect changes in file popularity, cosine similarity's relative insensitivity to small fluctuations makes it easier to set a consistent threshold.

\paragraph{Transfer Learning Adaptation and Comparison}

\begin{figure}
  \begin{center}
  \includegraphics[width=3in]{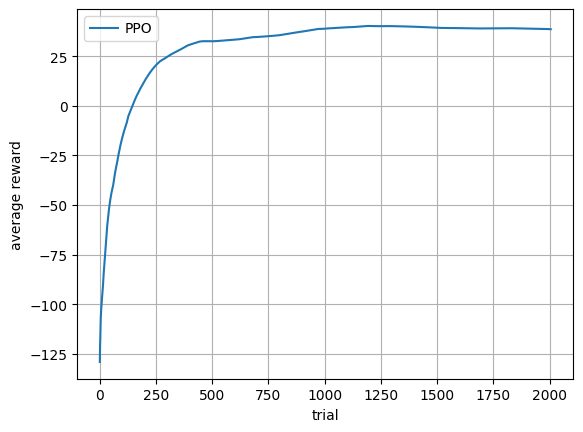}
  \caption{Convergence of the proposed caching approach in the initial environment}
  \label{source}
  \end{center}
\end{figure}

In this section, we consider two distinct types of changes in the environment to thoroughly evaluate the robustness and adaptability of the proposed algorithms. The first type focuses on a scenario where only the content popularity changes, keeping the overall request rate constant. The second type explores a scenario where content popularity remains unchanged but the request rate changes. Before diving into these dynamic scenarios, we first demonstrate the convergence of the caching approach in the initial environment, as illustrated in Figure \ref{source}.

\begin{figure}[!t]
    \centering
    \subfloat[]{\includegraphics[width=3in]{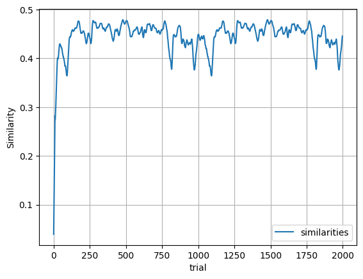}%
    \label{PLhit}}
    
    \hfil
    
    \subfloat[]{\includegraphics[width=3in]{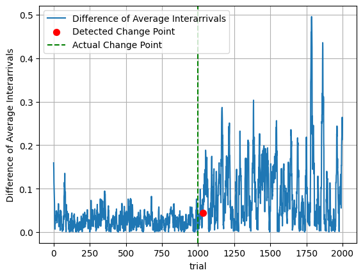}%
    \label{PIhit}}
    
    \caption{a) Change in Popularity Detection b) Change in Request Rate Detection}
\end{figure}

\begin{figure}
  \begin{center}
  \includegraphics[width=3in]{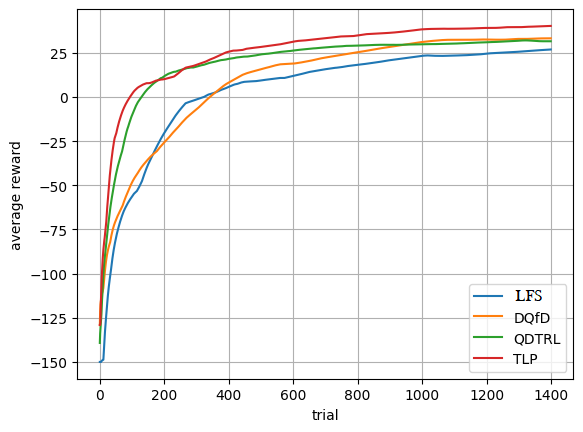}
  \caption{Average reward over time in the new environment under request rate change}
  \label{AvgRRC}
  \end{center}
\end{figure}

\begin{itemize}
  \item \textbf{Performance Comparison Under Popularity Change:}
  In the first experiment, the popularity of the most and second-most popular files is swapped with the least and second-least popular files, while maintaining the total request rate unchanged. This setting aims to evaluate how the algorithms respond to a shift in content popularity without altering the overall user demand pattern. Fig. \ref{CDP_Alpha} and \ref{CDP_Interarrival} present the outputs of our mechanism for popularity change detection and request rate change detection functions, respectively. As expected, the popularity change detection mechanism promptly identifies the swap, as the drastic reordering of popularities alters the request frequencies for these specific files. However, the request rate change detection mechanism correctly does not signal any changes, since the request rate remains unaltered. This confirms that the request rate detection mechanism is well-calibrated and does not produce false positives when only popularity changes are observed.
  
  \hspace{2mm} Figure \ref{AvgRPC} illustrates the average reward over time for four algorithms: LFS, DQfD, QDTRL, and TLP. LFS begins learning from scratch and lacks any prior knowledge to assist in the new environment. Without any information transfer, LFS has to explore the environment extensively before it can identify an optimal caching strategy, leading to a prolonged convergence period. The performance of QDTRL is comparable to that of LFS, showing only marginal improvements. This is likely because QDTRL still struggles when the changes are substantial, rendering much of the previously learned information less useful. This suggests that QDTRL might require more sophisticated adaptation mechanisms to effectively handle severe environmental changes and benefit from the prior knowledge. DQfD outperforms both LFS and QTDRL in terms of achieving faster convergence. The primary reason is its ability to effectively leverage expert demonstrations to guide early-stage learning, reducing the exploration time needed to identify optimal policies. On the other hand, TLP significantly outperforms all other baselines and converges much faster. TLP outperforms the other baselines primarily because of the way it prioritizes transitions, which enables it to demonstrate more stabilized behavior in response to popularity changes. By effectively recognizing and assigning higher importance to critical transitions, TLP adapts more efficiently to the new environment. Furthermore, TLP’s ability to relearn its action selection policy while leveraging knowledge from the source environment allows it to strike a balance between exploration and exploitation. This is particularly advantageous in scenarios where the popularity of files changes drastically, as it can exploit the most relevant transitions while exploring new patterns. Another key advantage of TLP is its gradual overwriting of demonstration data. Unlike DQfD and QDTRL, which suffer when the differences in popularity distributions between environments are too high, TLP can progressively phase out outdated information, making it robust to significant shifts in file popularity and ensuring improved performance even under abrupt changes. This makes TLP not only adaptable to the new environment but also highly effective in retaining critical knowledge while adjusting its policy accordingly.

  \begin{figure}[!t]
    \centering
    \subfloat[]{\includegraphics[width=3in]{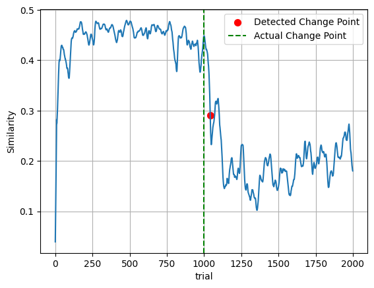}%
    \label{CDP_Alpha}}
    
    \hfil
    
    \subfloat[]{\includegraphics[width=3in]{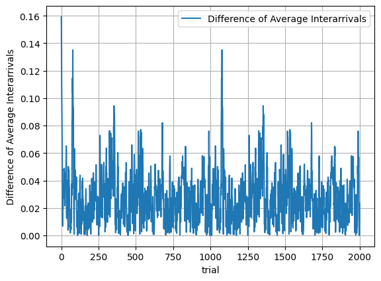}%
    \label{CDP_Interarrival}}
    
    \caption{a) Change in Popularity Detection b) Change in Request Rate Detection}
\end{figure}

\begin{figure}
  \begin{center}
  \includegraphics[width=3in]{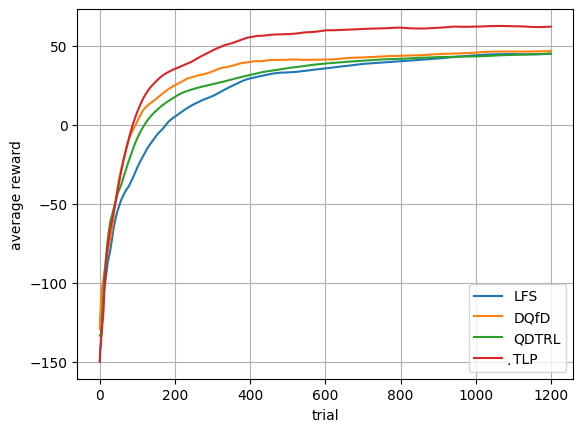}
  \caption{Average reward over time in the new environment under popularity change}
  \label{AvgRPC}
  \end{center}
\end{figure}

  \item \textbf{Performance Comparison Under Request Rate Change:}

  In this subsection, we consider a scenario where the request rate changes while the popularity distribution of the content remains constant. Befre change $\lambda = 5$, and after change $\lambda = 3.3$. Figure \ref{PLhit} and Figure \ref{PIhit} show the outputs of the popularity and request rate change detection functions, respectively. As expected, the popularity change detection algorithm does not detect any change, confirming that the content popularity remains stable. On the other hand, the request rate change detection algorithm promptly identifies the change in request rate, highlighting its sensitivity to variations in request patterns. 
  
  \hspace{2mm} Figure \ref{AvgRRC} illustrates the performance of TLP and baseline algorithms in the new environment after the change. LFS consistently performs the worst, as it lacks the ability to transfer knowledge from the previous environment and must start learning from scratch. DQfD is closely aligned with LFS, likely due to its reliance on outdated demonstration data that is not adapted to the new request rate pattern. This similarity arises because DQfD struggles to adapt its policy when the distribution of requests shifts, causing it to converge slowly in response to the change. QDTRL, in contrast, performs better than both LFS and DQfD, as it partially transfers knowledge from the previous environment. However, its improvement is limited because it does not fully account for request rate variations, making its adaptation slower. TLP significantly outperforms all other baselines and converges quickly in the new environment , for the same reason stated earlier. More importantly, TLP explicitly models the request rate using an SMDP formulation. This structured approach allows the algorithm to better recognize changes in the request rate, leading to more informed decision-making. By integrating SMDP, TLP achieves a balance between utilizing existing knowledge and exploring the new environment, allowing it to adjust efficiently and maintain a high average reward even when the request dynamics shift.

\end{itemize}

\section{conclusion} \label{conclusion}

In conclusion, this paper presents an effective edge caching strategy that addresses the challenges of dynamic environments by integrating key file attributes and accounting for random file request arrivals. By incorporating PPO and developing an effective mechanism for detecting changes in content popularity and request rates, the proposed solution adapts efficiently to variations in the environment, avoiding the computational costs of learning new policies from scratch. The transfer learning-based PPO algorithm further accelerates convergence by leveraging prior knowledge, leading to faster adaptation and improved performance. Simulation results confirm that our approach can promptly and effectively detect environment changes, and the proposed transfer learning approach significantly outperforms recent transfer learning-based methods in terms of convergence, demonstrating its potential to enhance edge caching in dynamic real-world scenarios.

\vfill

\end{document}